# INVESTMENT VOLATILITY: A CRITIQUE OF STANDARD BETA ESTIMATION AND A SIMPLE WAY FORWARD

Dr. Chris Tofallis
Department of Management Systems
The Business School
University of Hertfordshire
College Lane
Hatfield
AL10 9AB
United Kingdom
Email: C.Tofallis@herts.ac.uk

## Abstract

Beta is a widely used quantity in investment analysis. We review the common interpretations that are applied to beta in finance and show that the standard method of estimation – least squares regression – is inconsistent with these interpretations.
We present the case for an alternative beta estimator which is more appropriate, as well as being easier to understand and to calculate. Unlike regression, the line fit we propose treats both variables in the same way. Remarkably, it provides a slope that is precisely the ratio of the volatility of the investment's rate of return to the volatility of the market index rate of return (or the equivalent excess rates of returns). Hence, this line fitting method gives an alternative beta, which corresponds exactly to the relative volatility of an investment - which is one of the usual interpretations attached to beta.

## 1. Clearing up some basics

In the world of finance the term 'beta' refers to the slope in a linear relationship fitted to data on the rate of return on an investment and the rate of return of the market (or market index). This usage stems from Sharpe's 1963 paper in Management Science where he actually used the Roman letter B rather than the Greek β. (Strictly speaking, in statistics Roman letters refer to measured or estimated values based on a sample of data, whereas Greek symbols refer to the true, but unknown population values.)
The relationship is usually stated in one of two forms:

$$R_i = \alpha + \beta R_m \qquad (1)$$

Where $R_i$ represents the rate of return on an investment (e.g. in percentage terms), and $R_m$ is the rate of return on the market or an index of the market. As it stands, (1) is the equation of a line fitted to the data, with α and β being the intercept and slope of that line; an error term will be required when referring to particular data points.

It is well worth stressing that verbal explanations of beta are often incorrect and give the wrong impression. For example, the head of investment funds at Cazenove Fund Management in an article explaining various risk measures, makes the usual mistakes: "if a stock has a beta of 1.5 and the market rises by 1%, the stock would be expected to rise by 1.5%" (Minter-Kemp, 2003). This is wrong on two counts: firstly, it should be referring to a change in the *rate of return* of the market – not changes in the index itself, and secondly, it should refer to a change in the stock's rate of return, not in its price. Sadly, such careless wording sometimes appears in textbooks too (e.g. Hirschey, 2001, p.540). In fact, on a graph with $R_i$ on the vertical axis versus $R_m$ on the horizontal axis, if the market rises by 1% then this will merely refer to a single point on the graph and so there is no slope to be measured! To estimate beta one needs (at the very least) two data points. Each data point refers to rates of return over a time *interval, say t to t+1*. Hence to estimate the slope one needs measurements over at least two time *intervals,* say *t to t+1* and *t+1 to t+2,* which implies knowledge of stock and index prices at *three* points in time. The incorrect explanation gives the impression that only two points in time are needed to understand beta.

The other form of the linear relationship deals with 'excess returns' i.e. the rate of return above and beyond that which is available from a risk-free investment such as lending to the government:

$$R_i - r_f = \alpha + \beta (R_m - r_f) \qquad (2)$$

where $r_f$ is the rate of return of the risk-free asset. An excess return is sometimes called a 'risk premium'. The line associated with (2) is called the characteristic line for that investment.

If we re-plot our graph and replace the variables by *excess* rates of return, then each original point will have each of its coordinates reduced by $r_f$. However, this does not mean that all points will have been shifted by the same amount. This is because the risk-free rate is not always the same: when this rate changes, then subsequent data points will be shifted by a different amount. Consequently, estimates of beta from these two equations will not be identical. According to Bodie et al (2002, p304) most commercial providers of beta data do not use the excess return form.

## 2. Standard beta

The standard textbook way of estimating beta uses ordinary least squares (OLS) regression with the left hand side of (1) or (2) as the dependent variable. The resulting slope can be expressed as

$$\beta = r \, \sigma_i / \sigma_m \qquad (3)$$

where the $\sigma$'s are the standard deviations of the rates of return and r is the correlation between the rates of return. We shall refer to this as 'standard beta'. An equivalent formula is the ratio: (covariance between market and investment returns)/ (variance of the market returns).

This method of estimation makes the important assumption that the independent variable (market return) does not have any error associated with it. If one is using a market index as a proxy for the market (as in the capital asset market model, CAPM) then there will be error present. This is called the errors in variables problem or benchmark error. Note that simply moving from an index such as the Dow Jones Industrial Average (only 30 stocks) to a broader index such as the S&P500 hardly dents this problem since the 'market' in CAPM refers to the universe of all investments, which includes foreign equities, bonds, land, property, gold, derivatives, foreign currencies etc. In fact, it was part of Roll's (1977) famous critique of CAPM that it was not a testable theory unless we know the exact composition of the market portfolio. Whilst there are estimation methods for dealing with measurement error in the independent variable, they require knowledge about the variance of the error – and this is simply not known. What can be said however is that the resulting betas would have a higher value than standard beta. This under-estimation is true for the usual case of positive values of beta; if beta were negative then the measurement error estimator would be even more negative. Thus, in general, the correction arising from the benchmark error will move the beta estimates further away from zero.

Let us suppose that we are not using the market index as a proxy and that we are quite content to relate our returns with those of our chosen index as benchmark. Regression models minimize the sum of squared errors in the dependent variable only – this is because the purpose of regression is to fit a relationship for predicting the dependent variable (rate of return of the investment) for a stated value of the explanatory variable (the market rate of return). Statisticians might however be surprised to learn that betas are rarely used for such a purpose! It thus makes sense to survey the common uses of beta in finance and see if the least squares estimator is ever appropriate. We shall do this in the remainder of this paper and we will argue that the widely used least squares estimator is inappropriate.

## 3. Beta used to apportion risk to the market

In general, the linear relationship with the market returns (1) will not be perfect: most points will not lie on the line and so there is an error term (e) to consider:

$$R_i = \alpha + \beta R_m + e \qquad (4)$$

The term $\beta R_m$ is supposed to represent the part of the return which is explained by market variations, and the error term accounts for non-market variations.
This seemingly plausible decomposition is very likely untrue - we need to be more careful: We have made a huge assumption in thinking that the relationship between $R_i$ and $R_m$ is a nice straight line. If a non-linear relationship were fitted the error term would no doubt be lower, this is quite simply because nonlinear relations are obviously more flexible and can get closer to the data. As a result of the better fit the variation attributed to the market would then be higher and the remaining 'non-market' variation lower. Hence the relative attribution ('sharing out the risk') into market risk and investment-specific risk is highly dependent on the functional form of the underlying model that is chosen.

But that is not the only problem with this apportionment. Let us play along for a while longer and assume the relationship with market rate of return is truly linear. The argument for decomposition of risk into market risk (also known as systematic risk) and investment-specific risk (unsystematic risk) runs as follows. Let 'var' denote variance, then assuming the terms on the right hand side of (4) are uncorrelated, we have:

$$\text{var}(R_i) = \text{var}(\alpha) + \text{var}(\beta R_m) + \text{var}(e) \qquad (5a)$$

we are then told that "$\alpha$ and $\beta$ are constant" from which it follows that
$\text{var}(\alpha) = 0$, and
$\text{var}(\beta R_m) = \beta^2 \text{var}(R_m)$
hence $\quad \text{var}(R_i) = \beta^2 \text{var}(R_m) + \text{var}(e)$
$\quad\quad\quad\quad\quad = \text{market risk} + \text{investment-specific risk} \qquad (5b)$

This shows beta's role in apportioning risk. "For very well diversified portfolios, non-systematic risk tends to go to zero and the only relevant risk is systematic risk measured by beta" (Elton et al 2003). Thus the term containing beta is also called the non-diversifiable risk.

The trouble with the above argument lies in the assumptions: the fact is that beta (and therefore alpha) are *not* constant – this effectively destroys the above derivation. (For example Hirschey (2001, p.546) shows that for Dow Jones stocks the correlation between current year betas and previous year betas is only 0.34. Chawla (2001) reviews the literature on beta stability and uses hypothesis tests to demonstrate instability.) If betas were constant then we could look them up for any particular stock in some Eternal Beta Bible knowing that the value we found would be true for all time. In fact, it is precisely because they are changing that there is a demand for 'beta books' which is catered to by data providers such as Value Line Investment Survey, Bloomberg, Standard and Poor's, Ibbotson Associates and the Risk Measurement Service of the London Business School. The literature tells us of a tendency for standard betas values to approach the value of unity over time. As a result there have been attempts to capture this tendency. These include Blume's beta (a weighted average of standard beta and one) and Vasicek's beta (a weighted average of standard beta and the average beta for a sample of stocks). Shalit and Yitzhaki (2002) discuss the instability of OLS estimators of beta, and blame the quadratic loss function which makes extreme observations have a magnified effect. They propose the use of a coefficient to represent the investor's risk aversion. Martin and Simin (2003) also focus on the effect of such outliers, and observe that the effect is particularly noticeable for small firms. They recommend using a weighted least squares estimator where the weights are determined by the data. Other models which specifically aim to capture the time-variation of beta have been developed, see Faff et al (2000) for a comparison.

Fabozzi and Francis (1978) investigated 700 stocks on the New York stock exchange and found that "many stocks' betas move randomly through time rather than remain stable as the ordinary least squares model presumes". They demonstrate that the partitioning of risk "will be confounded with the noise from the shifting beta. As a result it will not be possible to estimate empirically the separate effects of systematic and unsystematic risk…this particular implication undermines too many empirical studies to list here".

In conclusion, the fact that beta values change means that the standard apportioning of risk into market risk and diversifiable risk as derived above (5a, 5b) is flawed, because the derivation assumes a constant beta.

## 4. Beta as relative volatility

We shall now show that the standard interpretation of beta is not consistent with the formula used to estimate it. This is extremely important because many financial decisions are being made daily by analysts using this interpretation.

Volatility is measured in the financial context by the standard deviation of the rates of return, and is often used as a measure of risk. Hence, if we wish to compare the volatility of an investment's rates of return with the volatility of the market rates of return then one would expect to simply use the ratio

$$\sigma_i / \sigma_m = \text{relative volatility or volatility ratio} \qquad (6)$$

Logical, yes, but disappointingly it is not this ratio, but rather formula (3) i.e. beta, that according to textbooks is supposed to give us the relative volatility: "Beta measures the volatility of a given asset relative to the volatility of the market" (Levy, 2002); "Beta measures how volatile a fund has been compared with a relevant benchmark" (Hirschey, 2001). Sharpe (the originator of this financial statistic) et al (1999, page 183), make the same interpretation: " Stocks with betas greater than one are more volatile than the market and are known as aggressive stocks. In contrast, stocks with betas less than one are less volatile than the market index and are known as defensive stocks". Yet, one look at equation (3) shows us that standard beta is not the same as relative volatility, (6). There is something inconsistent here. If an investment had the same risk (volatility, $\sigma_i$) as the market then its volatility ratio would equal unity, but standard beta would not equal unity. Instead, its beta value would, from (3), equal its correlation with market returns, and hence would always be less than unity. Hence, the usual classification into aggressive and defensive stocks falls apart if one is using these terms to refer to relative volatility. The formula for standard beta (3), confounds (mixes together) relative volatility and correlation. Therefore, a low beta could actually represent a high relative volatility that is being masked by a low correlation. Investors would then be mistaken in thinking that they had selected an investment whose volatility was low. For example take a look at Figure 1.

Figure 1 compares a monthly time series plot of AT&T's excess returns with those of the S&P500 Index over the same five-year period. From the graph, one can see that AT&T (a telecom stock) is more volatile than the index. Yet the beta value for AT&T over this period is actually 0.75, and since this is less than unity this statistic gives the impression that this stock is less volatile than the index. One can understand how this arises when one is informed that the correlation is only about 0.32. One can now deduce the relative volatility (6) as $\beta/r = 0.75/0.32 = 2.34$. This being in excess of one is in agreement with our intuition when looking at the graphs. On repeating the analysis with the 30 stocks making up the Dow Jones Industrial Average, one finds that half of them had standard betas less than unity. Since any index is essentially a weighted average of its components, basic statistics tells us that we would expect it to be less variable than its components (central limit theorem), not more so. It is strange that analysts accept unquestioningly claims that so many stocks are less volatile than the market as a whole.

Camp and Eubank (1985), observed that many investors do not hold well-diversified portfolios, and so for them market risk is an incomplete risk measure. So they suggested use of the ratio of standard deviations (6) – which they called 'beta quotient' – as a measure of risk. "Because beta fails to consider unsystematic/diversifiable risk…the authors propose a risk measure that takes into account total variation of return relative to overall market variation". "The return performance of a portfolio should be evaluated on the basis of its beta quotient instead of its beta, since it is bearing diversifiable risk in addition to its systematic or non-diversifiable risk".

## 5. Beta in CAPM

The security market line is a linear relation that is fitted to data on average excess returns of a number of assets (dependent variable) and their standard beta values (explanatory variable). Since beta is here being used as a measure of risk, there is an expectation that higher beta stocks will have higher returns. The parameter values (slope and intercept) of this fitted line have been used to test the CAPM theory. A famous study by Fama and French (1992) showed that the slope was not significantly different from zero i.e. there was no positive association between return and standard beta. However there are other researchers who disagree with these findings. Roll and Ross (1992) claim that the choice of market index that is used to estimate beta can affect such conclusions. This is the errors-in-variables problem: since there is error in our measurement of the "market" return, this will affect the estimate of the slope (beta). OLS only assumes error in the dependent variable.

One can prove (e.g. see Elton et al, 2003, p 358) that if the explanatory variable has a random error and even if the mean of the errors is zero, this will still lead to a slope estimate in the security market line which is too low (downward biased). This in turn implies that the estimate for the intercept will be too high.

It would therefore seem desirable to: (i) estimate beta in a way that allowed for measurement error in the variable which is chosen as a proxy for market return, and (ii) estimate the security market line in a way which allowed for error in the explanatory variable

## 6. Alpha as a risk-adjusted performance measure

Betas often play a part in the construction of risk-adjusted measures of performance. These measures are subsequently used for ranking the desirability of investments. The idea is that if two investments have the same total returns, we should prefer the one that has been less volatile. One sometimes sees discussions in the financial press that mention a fund manager's alpha. This is not a part of their anatomy. It is used as a measure of performance that takes into account the level of risk (as measured by beta) that has been taken. To see this, take a look at equation (1): the return produced by an investment is split into two parts. One part ( $\beta R_m$ ) shows the return attributable to market changes for the level of risk ($\beta$) taken on. The other term ($\alpha$) is unrelated to market movements and is interpreted as being the return attributable to the fund manager's skill (or luck). Hence positive alpha is often used as a hallmark for investor talent. For a given set of data, the way we estimate $\beta$ will have an effect on the consequent value of $\alpha$: if we under-estimate beta, then we shall over-estimate alpha. If the arguments in the next section are to be believed then that is precisely what has been done in the past: beta (risk) has been underestimated, and consequently the skill of fund managers has been over-estimated. This is not something that applies uniformly to all investment managers i.e. their alpha scores will not merely be shifted such that their rankings stay unchanged, rather, the new alphas will rank managers in a different order. [1]

## 7. A way forward

We have looked at various roles that beta has been given and found that the standard method of estimating beta has shortcomings. Let us return to the beginning and see if we can do things differently. We start with a set of points on our graph with investment rates of return on the y-axis and market rates of return on the x-axis. The following arguments are unaltered if excess rates of return are used. We want to plot a straight line and estimate the slope of this line. Previously we used ordinary least squares (OLS) regression. But wait, there are two regression lines! The OLS line minimizes the sum of squared deviations in the y-direction. The reverse regression line minimizes in the x-direction. If our purpose is predicting y for a specified x-value, statisticians will advise use of OLS regression. If our purpose is to predict x for a specified y-value we are advised to use reverse regression. However none of the usual interpretations for beta that we have discussed includes either of these purposes. What we in fact require is the *slope* of the *functional relationship* between x and y. As Kendall and Stuart emphasise in their classic statistics text (1979, p 402): "A regression line does not purport to represent a functional relation between mathematical variables or a structural relation between random variables". Many practitioners and researchers – even statisticians – often forget this; they inadvertently slip into thinking that their OLS model estimates the underlying relationship between variables. This probably arises because methods for fitting functional relations do not usually appear in current statistics textbooks, and so students are not aware of the fact that there are other ways of fitting lines to data.

One basic fact from statistical theory is that the slopes of the two least squares regression lines bracket the slope of the estimated functional line. This is to be expected since the ordinary regression line is estimated by minimising all the variation in one direction and the reverse regression minimises all the variation in the other. Booth and Smith (1985) therefore suggested using the two regression estimates as bounds on the true value.

We now have upper and lower limits for the slope but which value shall we settle upon? A sensible approach is to choose one that carries with it those roles that beta has been used for in the past that have not been put into question. Let us consider the relative volatility role (volatility relative to the market). We said earlier that a more logical estimator for this purpose would be the ratio of standard deviations (6). Since this is always positive we need to attach a sign. This will be given by the correlation; this ensures that we can also deal with downward sloping characteristic lines. We now investigate this alternative estimator of beta, denoting it by $\beta^*$.

$$\beta^* = (\text{sign of } r)\ \sigma_i / \sigma_m \qquad (7)$$

---

[1] I am grateful to one of the referees for pointing out that non-parametric approaches to risk-adjusted performance measurement are a currently active area of research. See Galagedera (2004) for a survey.

or the equivalent form which uses the standard deviations of the excess rates of return. The connection with the standard OLS beta is apparent from (3):

$$\beta^* = \beta / r \qquad (8)$$

Does this estimator lie between the two regression slopes as required? The reverse regression slope is given by $\beta / r^2$. (Incidentally, this shows how large the differences in regression estimates can be: a correlation of 0.71 implies that reverse regression has a slope twice as high as the standard regression!) Since $\beta^*$ equates to $\beta/r$ and since r lies between –1 and +1 it follows that our proposed estimator does satisfy the requirement of lying in between. For the usual case of positive correlation between market and the investment, we have the standard beta giving the lowest value and the reverse regression the highest, so we have:

$$\beta \leq \beta^* \leq \beta_{reverse}. \qquad (9)$$

The equalities hold only when there is perfect correlation in the data. This is as one would expect, as then all points lie exactly on a straight line and so there can be no disagreement on where the line should be.

Does this new slope estimator correspond to an established line fitting procedure? In fact it does: it is precisely the geometric mean functional relationship (Draper and Smith, 1998). Its name refers to the fact that the slope is the geometric mean of the slopes from the two least squares regressions: i.e. multiply those slopes and take the square root. This also implies that its value lies between the ordinary and reverse regression slopes. This line also passes through the centroid of the data i.e. the point whose coordinates are the mean values of the plotted variables. This is the only point which all three lines pass through.

Another point in favour of our estimator is its symmetric functional form. If we had only two data points we would estimate the slope as "(rise in y)/(rise in x)"; notice that this treats changes in the y-variable in the same way as changes in the x-variable. The volatility ratio, equation (6), maintains this symmetry in the treatment of the two variables. However the equation for standard beta (3) does not – one need only inspect the formula for correlation to see this.

Is this line optimal in any way? Yes it is, and what is more it is optimal in a way that involves both the vertical and horizontal deviations from the line. In fact it minimizes the sum of products of these deviations. This is equivalent to saying that it is the line that minimizes the sum of the areas of the triangles made by the points and the line (see Figure 2). This was proved by Woolley (1941). From this it follows that the estimated relationship between the two variables will be the same irrespective of which variable is plotted on each axis i.e. there is symmetry of treatment: each variable is treated with equal importance. This is just how we would want to treat variables if we were aiming to discover an underlying relationship between them.

To compare values of the proposed estimator with standard beta refer to Table 1. Notice how, as well as the new values being higher, the relative risk rankings are also now quite different.

Draper and Smith (1998 p.92) have started to promote the use of this line in the latest edition of their book on regression, but are unaware that one can also establish relevant confidence intervals. Kermack and Haldane (1950) demonstrated that the formula for the variance of our estimator can be approximated by that for the OLS case,
i.e. the variance of the slope is
$$s^2 = \beta^* (1-r^2) / (n-2) \qquad (10)$$
where n is the number of data points.
A confidence interval can be constructed in the usual way using the Student t-distribution: $\beta^* \pm t\,s$
An exact form for the confidence interval due to Jolicoeur and Mosimann is given in Ricker (1984), namely:
$$\beta^* [\,(B+1)^{1/2} \pm B^{1/2}\,] \qquad (11)$$
where $B = t^2 (1-r^2) / (n-2)$.

What can we say about the stability of the proposed beta estimator? Francis (1979) looked at stability from the point of view of the different parts of the formula for standard beta (see equation (3)). He found "explicit evidence pinpointing each stock's correlation with the market as the most unstable statistic within beta". His conclusion is that "the correlation with the market is the primary cause of changing betas…the standard deviations of individual assets are fairly stable". This bodes very well for our estimator since it differs from standard beta in precisely not including the correlation between the investment and the market. Hence we expect it to be more stable over time. As a small test we looked at stocks in the Dow Jones Industrial Average calculating their standard betas for the period 1989-1994 and comparing them with those of 1995-2000. The absolute percentage change ranged from 1% to 100%, with a mean change of 23%. When this comparison was done using $\beta^*$, the change ranged from 0.3% to 45% with a mean change of only 15.7%. So we have some preliminary evidence that $\beta^*$ is more stable in time.

## 8. Conclusion

A key message of this paper is that OLS regression lines are not intended to represent an underlying relationship between two variables. Sadly, this misconception is one that is widespread. Rather, regression lines are intended for predicting the value of a dependent variable for a given value of an explanatory variable. If you switch the variables in an OLS regression you produce a different line, and so you don't have a unique relationship. This confusion between functional relationships and regressions can be traced back to Sharpe's seminal 1964 paper. When speaking of a plot of the rate of return on an asset ($R_i$) versus the rate of return on an efficient 'combination' of assets (the market portfolio), he says (p.438): "Part of the scatter of $R_i$ is due to an *underlying relationship* with the return on the combination, shown by B, the slope of the *regression* line". [Our italics.]

In an effort at estimating a unique underlying relationship, we therefore proposed a fitting technique which treated both variables on an equal footing. The resulting line is variously referred to in statistics as the geometric mean functional relation or the reduced major axis. It is optimal in the sense that it is a 'least areas line', see Figure 2.

The magnitude of its slope, $\beta^*$, is precisely the ratio of volatilities (standard deviations) and so we can now accurately refer to it as 'relative volatility'. This slope value lies between the slope values arising from ordinary regression and reverse regression. The only difference between its calculation and that from OLS is that its formula does not contain the correlation. Since it is the correlation that has been found to be the main contributor to instability in betas (Francis, 1979) we expect that $\beta^*$ will be more stable over time, and indeed we gave some preliminary evidence for this. Furthermore, the removal of the correlation from the formula brings clarity to what is being measured – there is no longer the confounding of two quantities: relative volatility and correlation. There is also a computational advantage in that it is easier to calculate the ratio of standard deviations than the OLS slope.

Our estimator is a measure of *total* risk and so it can be applied to all portfolios - whether they are diversified or not. A consequence of this, of course, is that it cannot play a part in splitting up risk into components (market risk and investment-specific risk). It must be stressed however that standard beta's claim to measure market risk is highly questionable – as we demonstrated the difficulty is primarily due to the instability of beta over time. Fabozzi and Francis, (1978) make this point most emphatically:

> "After Markowitz and Sharpe suggested estimating the beta systematic risk coefficient for market assets, finance professors, stock brokers, investment managers, and others began expending large quantities of resources each year on estimating betas. Unfortunately however, it appears that the ordinary least squares regressions used in nearly every instance may be inappropriate".

For any given data set the absolute value of our proposed estimator $\beta^*$ will be higher than that of standard $\beta$. From this it follows that alpha values will be revised downwards (since the line will always pass through the centroid point-which can be viewed as a fixed point of rotation). An important implication is that if the new alpha is used to rate investment managers or funds then there will be fewer of them with the much sought after positive alpha.

Very importantly, the proposed estimator for beta finally allows for consistency between its standard interpretation (as relative volatility) and the formula used for its calculation. This gives an alternative, and we would argue a more logical classification of stocks as being either aggressive or defensive. One dreads to think of the fortunes that have been invested on the basis that beta values were interpreted as meaning investments were less volatile than the market when in fact they were nothing of the sort.

We end with a few wise words of advice:

> Before deciding what straight line to use, you must decide what you want it for. Do you wish to estimate (predict) one quantity from another, or do you want a descriptive trend line relating two sets of observations.
> (Ricker, 1984)

In the light of this we need to critically review past research as well as current decision-making which is based on inappropriate statistical analysis because:

> OLS continues to be by far the most frequently used method even when it is obviously inappropriate. As a result, hundreds if not thousands of regression lines with too-small slopes are being published annually.
> (Riggs et al, 1978).

## Acknowledgments

I am grateful to Professor Mick Broadbent and the anonymous referees for taking the time to provide comments on earlier drafts of this paper.

Figure 1
*Relative volatility: The dashed line shows the excess returns of the S&P500 index over a period of five years. The full line shows the excess returns of AT&T, which is clearly more volatile, yet its beta is only 0.75 , a value which gives the impression that it is less risky than the index.*
*Data: 60 months ending January 2000.*

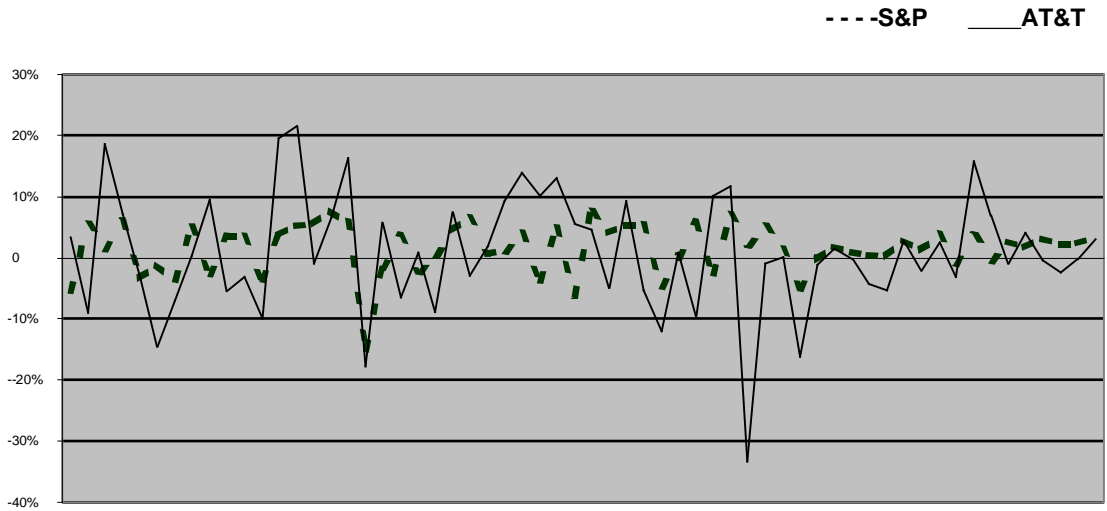

| Company | β | β* | β* Rank | β rank | Difference in ranks |
|---|---|---|---|---|---|
| Intel | 1.08 | 2.78 | 1 | 12 | 11 |
| Hewlett-Packard Co. | 1.28 | 2.69 | 2 | 5 | 3 |
| Alcoa Inc. | 1.13 | 2.58 | 3 | 10 | 7 |
| Microsoft | 1.45 | 2.56 | 4 | 3 | -1 |
| Citigroup Inc. | 1.67 | 2.37 | 5 | 1 | -4 |
| AT&T Corp. | 0.75 | 2.37 | 6 | 26 | 20 |
| IBM | 1.03 | 2.20 | 7 | 14 | 7 |
| Caterpillar Inc. | 0.88 | 2.15 | 8 | 19 | 11 |
| Walmart | 1.15 | 2.14 | 9 | 8 | -1 |
| International Paper Co. | 1.10 | 2.12 | 10 | 11 | 1 |
| Philip Morris Cos. Inc. | 0.55 | 2.04 | 11 | 28 | 17 |
| Home Depot Inc. | 0.97 | 2.02 | 12 | 15 | 3 |
| Coca-Cola Co. | 1.07 | 2.01 | 13 | 13 | 0 |
| Merck & Co. Inc. | 0.86 | 1.97 | 14 | 20 | 6 |
| United Technologies | 1.49 | 1.97 | 15 | 2 | -13 |
| Boeing Co. | 0.89 | 1.91 | 16 | 18 | 2 |
| Disney | 0.78 | 1.91 | 17 | 25 | 8 |
| Honeywell International | 1.14 | 1.89 | 18 | 9 | -9 |
| General Motors Corp. | 0.93 | 1.85 | 19 | 17 | -2 |
| American Express Co. | 1.34 | 1.83 | 20 | 4 | -16 |
| Du Pont de Nemours | 0.83 | 1.80 | 21 | 22 | 1 |
| J.P. Morgan Chase & Co. | 1.15 | 1.74 | 22 | 7 | -15 |
| SBC Communications | 0.82 | 1.72 | 23 | 23 | 0 |
| 3M | 0.62 | 1.66 | 24 | 27 | 3 |
| Johnson & Johnson | 0.96 | 1.64 | 25 | 16 | -9 |
| McDonald's Corp. | 0.81 | 1.64 | 26 | 24 | -2 |
| General Electric Co. | 1.22 | 1.62 | 27 | 6 | -21 |
| Eastman Kodak Co. | 0.29 | 1.59 | 28 | 30 | 2 |
| Procter & Gamble Co. | 0.84 | 1.56 | 29 | 21 | -8 |
| Exxon Mobil Corp. | 0.50 | 1.11 | 30 | 29 | -1 |

Table 1:
*If we rank the 30 companies making up the Dow Jones Industrial Average according to the proposed estimator and then according to standard beta, we observe large differences. Notice how technology companies Intel, IBM and AT&T now appear relatively much riskier than their standard betas would have led us to believe. Calculations based on 60 months ending January 2000.*

Figure 2: *The geometric mean functional relation is the line that minimizes the sum of the triangular areas defined by the points and the line. The slope ($\beta^*$) of such a line is precisely the ratio $\sigma_y/\sigma_x$ and since the line passes through the centroid of the data it follows that the equation of the line can be written as*: $y - \bar{y} = \beta^*(x - \bar{x})$. *For more details as well as extensions to multiple variables see Tofallis (2002, 2003).*

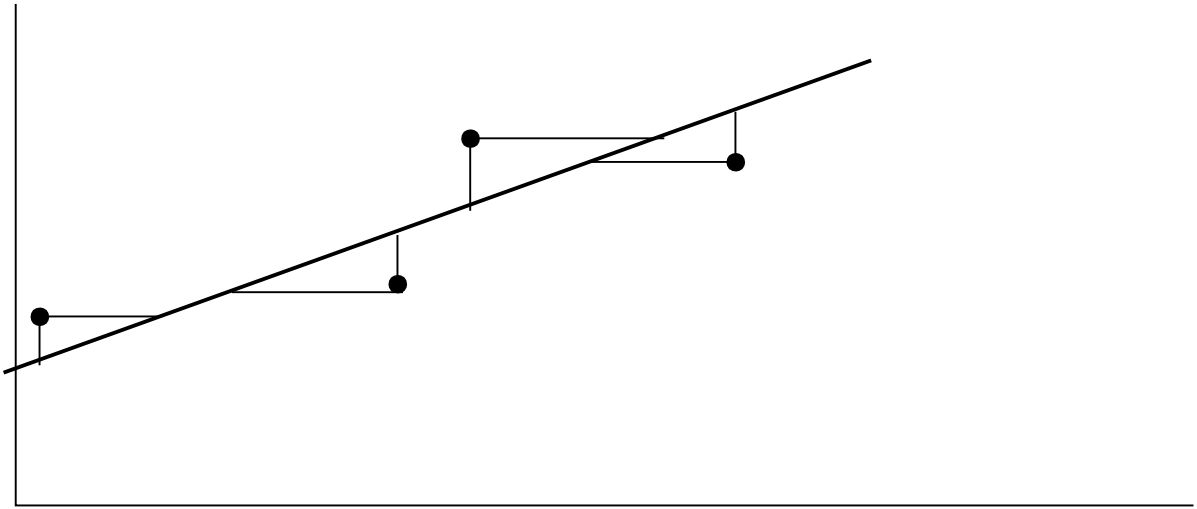